\input{aipcheck}

\documentclass[
    ,final            
  ]
  {aipproc}

\layoutstyle{6x9}

\begin{document}

\title{Emission Lines in X-ray Spectra of Clusters of Galaxies}



\author{Paolo Tozzi}{
  address={INAF -- Osservatorio Astronomico di Trieste -- via
G.B. Tiepolo 11 -- 34143 Trieste -- ITALY} }

\begin{abstract}
Emission lines in X--ray spectra of clusters of galaxies reveal the
presence of heavy elements in the diffuse hot plasma (the Intra
Cluster Medium, or ICM) in virial equilibrium in the dark matter
potential well.  The relatively simple physical state of the ICM
allows us to estimate, with good accuracy, its thermodynamical
properties and chemical abundances.  These measures put strong
constraints on the interaction processes between the galaxies and the
surrounding medium, and have significant impact on models of galaxy
formation as well.  This field is rapidly evolving thanks to the
X--ray satellites {\sl Chandra} and {\sl XMM--Newton}.  Among the most
relevant progresses in the last years, we briefly discuss the nature
of cool cores and the measure of the Iron abundance in high redshift
clusters.  Future X--ray missions with bolometers promise to provide a
substantial step forward to a more comprehensive understanding of the
complex physics of the ICM.
\end{abstract}

\maketitle

\section{Continuum X--ray emission from Clusters of Galaxies}

The 0.5-10 keV X-ray band is currently explored by the two X--ray
satellites {\sl Chandra} and {\sl XMM--Newton}, both launched in 1999.
Energies $E > 0.5 $ keV correspond to temperatures $T > 5 \times 10^6$
K.  Therefore X--ray astronomy is an ideal observational windows for
very hot astrophysical plasmas.

\begin{figure}
  \caption{Left panel: the X--ray emission (in red) of the cluster
    RXJ1252 at $z=1.235$ (observed with the {\sl Chandra} satellite)
    on top of the optical image (from the VLT).  Right panel: {\sl
      Chandra} X--ray spectrum of the extended emission in RXJ1252
    \cite{ros04}.  The Fe line clearly stands out at the rest--frame
    energy $E\sim 6.7$ keV.  }
\includegraphics[height=6cm]{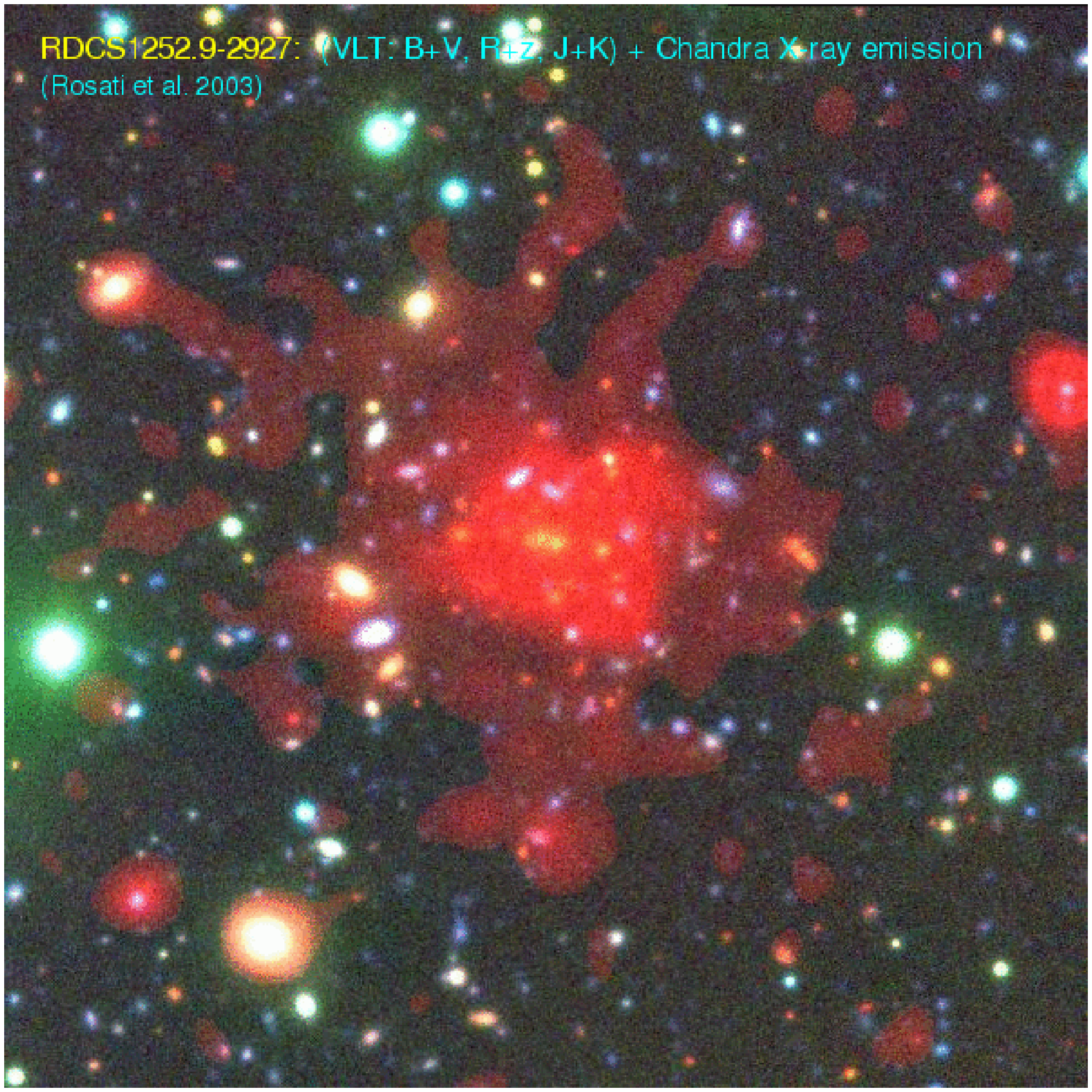} 
\includegraphics[height=6cm]{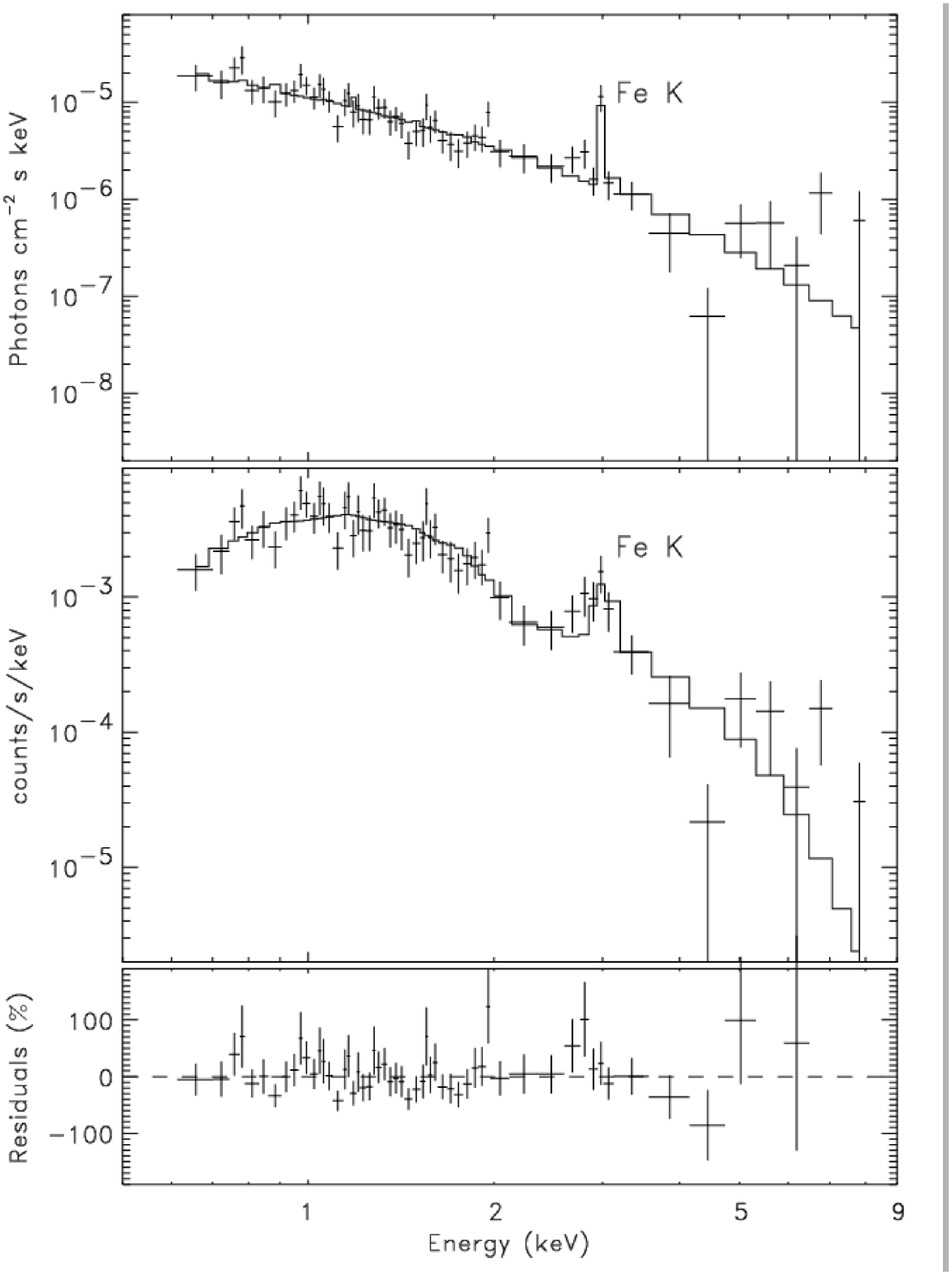} 
     \label{fig1}
\end{figure}

At the beginning of the era of X--ray astronomy, R. Giacconi and
collaborators detected X--ray emission from clusters of galaxies with
the Uhuru satellite.  This emission was identified as bremsstrahlung
from hot diffuse plasma (the Intra Cluster Medium, from now on ICM),
thanks to its extended distribution and the detection of higly ionized
Fe lines in the X--ray spectra.  Nowaday the imaging of bright X--ray
clusters of galaxies is possible up to redshifts as high as $z\sim
1.3$.  As an example, in Figure \ref{fig1} we show the diffuse X--ray
emission (in red) from the massive cluster RXJ1252 at $ z \sim 1.235$.
The ICM is generally smoothly distributed according to a {\sl 
beta}--profile, consisting in a flat core (sometimes hosting a
pronounced spike in emission as in the case of cool--core clusters)
followed by an almost isothermal slope $\rho_{ICM} \propto r^{-2}$
\cite{cff76}.  Typical total luminosities are observed to be roughly
in the range $L_X \sim 10^{43} - few \times 10^{45}$ erg/s.

Clusters are bound objects formed via gravitational instability from
the fluctuations in the primordial density field \cite{peeb}.  The
temperature of the diffuse baryons constituting the ICM is determined
by the energy equipartition between the mass components: dark matter,
($\sim 80$ \%), diffuse baryons ($\sim 17$\%) and stars in galaxies
($\sim 3$ \%).  The equipartition is reached through violent
relaxation during the gravitational collapse leading to the formation
of the cluster.  The dark matter, dynamicaly dominant, and the
baryons, rapidly adjust and reach a pressure balance with the
gravitational forces.  The velocities of particles become randomised,
and the structure settles into the {\sl virial} equilibrium.  In its
simplest form, the virial theorem writes as $2T+U = 0$, where $T$ is
the average kinetic energy per particle and $U$ is the average
potential energy.  This implies that, when virialization can be
applied, a measure of the temperature of the ICM gives a good estimate
of the total mass of the cluster.  Typical cluster masses are in the
range $10^{14} - 10^{15} M_\odot$, corresponding to $T_{ICM} \sim
1-10$ keV.

The continuum emission from the plasma is almost entirely given by
bremsstrahlung due to the acceleration of electrons in the Coulomb
field of positive ions (mainly Hydrogen and Helium nuclei).  The
relevant case for clusters of galaxies is bremsstrahlung emission
averaged over a thermal distribution of electron speeds.  This
emission can be computed with a classical treatment \cite{ryb} plus
quantum effects (included in the Gaunt factor of order unity) while
the relativistic corrections are often ignored (being few percent for
$kT_{ICM} > 10$ keV).  The thermal bremsstrahlung emission simply
scales as:

\begin{equation}
{{dE}\over{dt dV}} \propto n_e^2 T_{ICM}^{1/2}  \, .
\end{equation}

\noindent
This implies that the continuum emission of the ICM is strongly
dependent on its density distribution (through the electron density
$n_e$) but only weakly on its temperature.

\section{X--ray line emission from Clusters of Galaxies}

While the continuum emission is relatively simple to model, the total
emission must also include lines associated to the ions of heavy
elements distributed in the diffuse ICM.  First, it is important to
understand the kind of equilibrium which applies to the
ICM\footnote{The following description follows Kahn 2005\citep{k05}.}.
In general, equilibrium is given by a balance between competing
processes.  In the case of strict thermodynamic equilibrium, every
atomic process is as frequent as its inverse process (principle of
detailed balance).  The classic example is the black body.  
The detailed balance does not apply to the ICM, nevertheless, it
appears to be in the so called {\sl collisional} (or coronal)
equilibrium.  To understand what is the collisional equilibrium, we
first recall the three main ``actors'':

\begin{itemize}
\item  the kinetic distribution of electrons and ions;
\item  the atomic level populations;
\item the radiation field.
\end{itemize}

\noindent
Since the radiation field does not participate to the equilibrium, we
consider only the equilibrium between the electron and ion population
and the atomic levels.  The four key electron--ion collisional
processes (and their inverse) are:

\begin{itemize}
\item Collisional excitation (Collisional deexcitation);
\item Collisional ionization (Three body recombination);
\item Radiative recombination (Photoionization);
\item Dielectronic Capture (Autoionization).
\end{itemize}

Only few of these processes are relevant in the collisional ionization
equilibrium.  In fact, thanks to the very low density of the ICM,
atoms are always in their ground state, and collisional deexcitations
and three body recombinations are negligible.  Since the medium is
optically thin, photoionization, photo excitation and scattering are
neglected, so that excitation and ionization are dominated by electron
ion collisions.  Deexcitation is dominated by spontaneous radiative
decay, and ions will recombine by radiative or dielectronic
recombination.  It follows that the equilibrium is given by the
balance between impact ionization (and excitation autoionization) and
radiative and dielectronic recombination.  The interpretation of the
resulting spectrum will require a detailed knowledge of the
ionization, recombination and excitation rates of several transitions,
in particular of K--shell transitions of C, N, O, Ne, Mg, Si and Fe,
and L--shell transitions of Si, S, Ar, Ca, Ni and Fe.

To compute the collisional excitation rates we also assumes a
Maxwellian energy distribution of the electron energies, and a cooling
time longer than relaxation time (necessary to achieve equilibrium,
condition satisfied thanks to the low density of the ICM except
sometimes in the very inner regions).  Thus in a steady state the rate
of change of the population density $n_j$ of the $j$th ionization
state of a given element is given by:

\begin{equation}
0={{dn_j} \over {dt}} = S_{j-1}n_{j-1} - S_{j}n_j - \alpha_j \, n_J +
\alpha_{j+1} \, n_{j+1} \, ,
\end{equation}

\noindent
where $S_j$ is the ionization rate for ion j with ejection of one
electrons (direct ionization and autoionization), while $\alpha_j$ is
the recombination rate of ion j (radiative and dielectronic).  The
ionization structure is derived by solving for each element with
atomic number Z a set of Z+1 coupled rate equations.  In the steady
state the equation reduces to:

\begin{equation}
{{n_{j+1}} \over {n_j}} =  {{S_j(T_{ICM})} \over {\alpha_{j+1}(T_{ICM})  }} \, .
\end{equation}

Thus the ratio of two adiacent ionization states of a given element
depends only on $T_{ICM}$ and not on the electron density $n_e$ as
long as stepwise ionization (more than one collision) in $S_j$ and
three body recombination in $\alpha_{j+1}$ can be neglected.  The
fraction of ions at the stage z of a given element, $\eta_z$, can be
expressed directly as a function of the ratio of adiacent states.


\begin{figure}
\includegraphics[height=6cm]{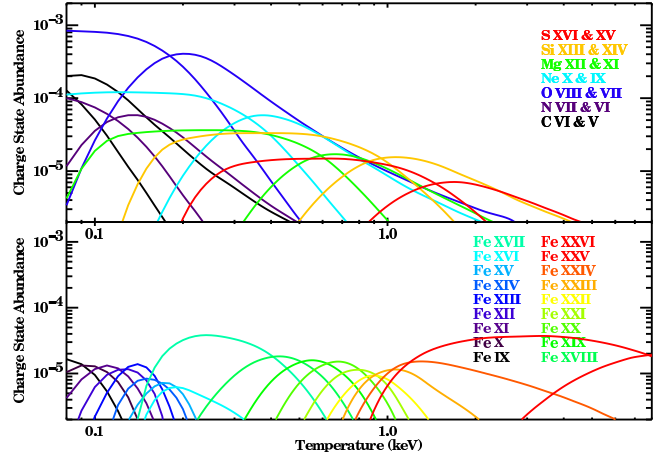} 
\caption{Upper panel: ion concentration of some elements as a function
  of $T_{ICM}$.  Lower panel: Fe ions concentration as a function of
  the ICM temperature (from \cite{pef06}).} \label{fig2}
\end{figure}

The emissivity of a given emission line (which is proportional to the
Equivalent Width, $EW$, of a line) is given by $\epsilon_{21} = n_e
n_i \gamma_{21}(T_{ICM}) E_{12}$ where $\gamma$ is the collisional
excitation rate for transition $E_{12}$ and $ n_i = A_Z
\eta_z(T_{ICM}) n_H$ is the density of the ion in the ground state.
Once the coefficient $\alpha$ and $S$ are known from fundamental
atomic physics, the concentration of each ion of a given element can
be straightforwardly computed from the measure of the equivalent
width.  An example of the ion concentration of some elements as a
function of $T_{ICM}$ is shown in Figure \ref{fig2}.  The equivalent
width can be measured directly through X--ray spectroscopy and it is
defined as:
 
\begin{equation} 
EW \equiv \int ( {{I_\nu -I_\nu^0} \over { I_\nu^0}}) d h\nu \, ,
\end{equation}

\noindent
where $I_\nu$ is the spectrum, $I_0$ is the continuum component, and
the integral is over an energy range close to the line.  Since $EW$ is
directly proportional to the ion concentration $n_i/n_H$, it depends
only on temperature and abundance. 

There are of course sources of uncertainties. The first one is due to
the accuracy of atomic physics.  Both ionization and recombination
rates can be uncertain by a factor 2-4.  Fortunately, ionization and
recombination rates for H--like and He--like ions, which emit lines
that are among the strongest in astrophysical plasmas, are known more
accurately.

The second critical aspect is given by the spectral resolution.
X--ray spectra may be obtained directly with CCD imagers or with
reflection gratings spectrometers.  In the first case, the spectral
resolution is limited and several lines often blend with each other.
Only few bright sources can be analyzed with grating spectroscopy, due
to the high S/N required.  Therefore, in some cases, the abundance of
several elements cannot be recovered accurately.

Finally, a source of uncertainty comes from an aspect intrinsic to the
physics of clusters: in general the ICM may have different
temperatures at different radii, with significant gradients in the
central regions.  This aspect can be modelled, or can be mitigated in
bright clusters thanks to spatially resolved spectroscopy and
deprojection techniques.  In any case, the presence of different
temperatures along the line of sight significantly affects abundance
measures.

\section{Iron abundance at high redshift}

Temperatures and abundances are measured at the same time (from line
ratios and shape of the continuum) with the use of fitting packages
which implement atomic physics and the collisional equilibrium
equations.  Heavy elements (generally called metals), among which Iron
is the most prominent, are always found in the ICM.  Over a wide range
of $T_{ICM}$, the EW of the Fe K--shell line (mostly from Fe + 24 and
Fe +25) is several orders of magnitudes larger than any other spectral
feature.  At lower energies O, Si, S, and L--shell transition in lower
Fe ions show significant EW.  Their abundance is consistent with being
produced by the elliptical cluster galaxies \cite{mv88}.  In general,
Fe abundance is observed to be about $Z \sim 0.4 Z_\odot$ (where the
Iron solar abundance is \cite{A05}) at least for $T_{ICM}> 5$ keV.
Abundances of other elements can be measured either in very high S/N
spectra, or in low temperature clusters, since for $T_{ICM} \leq 1$
keV the number of ions species of elements lighter than Iron become
significant (see Figure \ref{fig2}).  The ratio of the abundance of
the $\alpha$ elements over Fe is very relevant to understand the
relative contribution of TypeII and TypeIa SNe.  However, here we will
discuss only some results on the Fe abundance at high redshift.

\begin{figure}
\includegraphics[height=6.5cm]{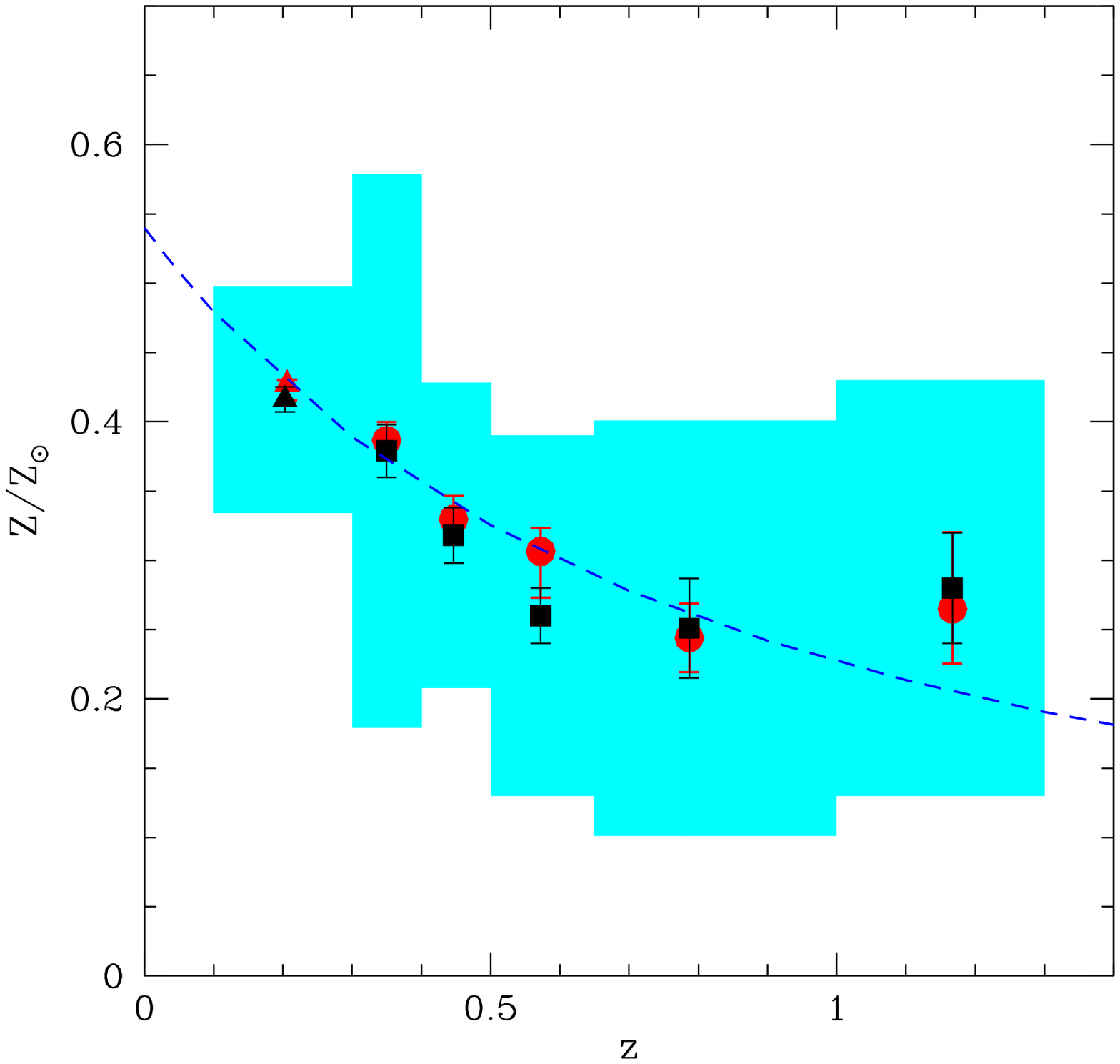} 
\includegraphics[height=6.5cm]{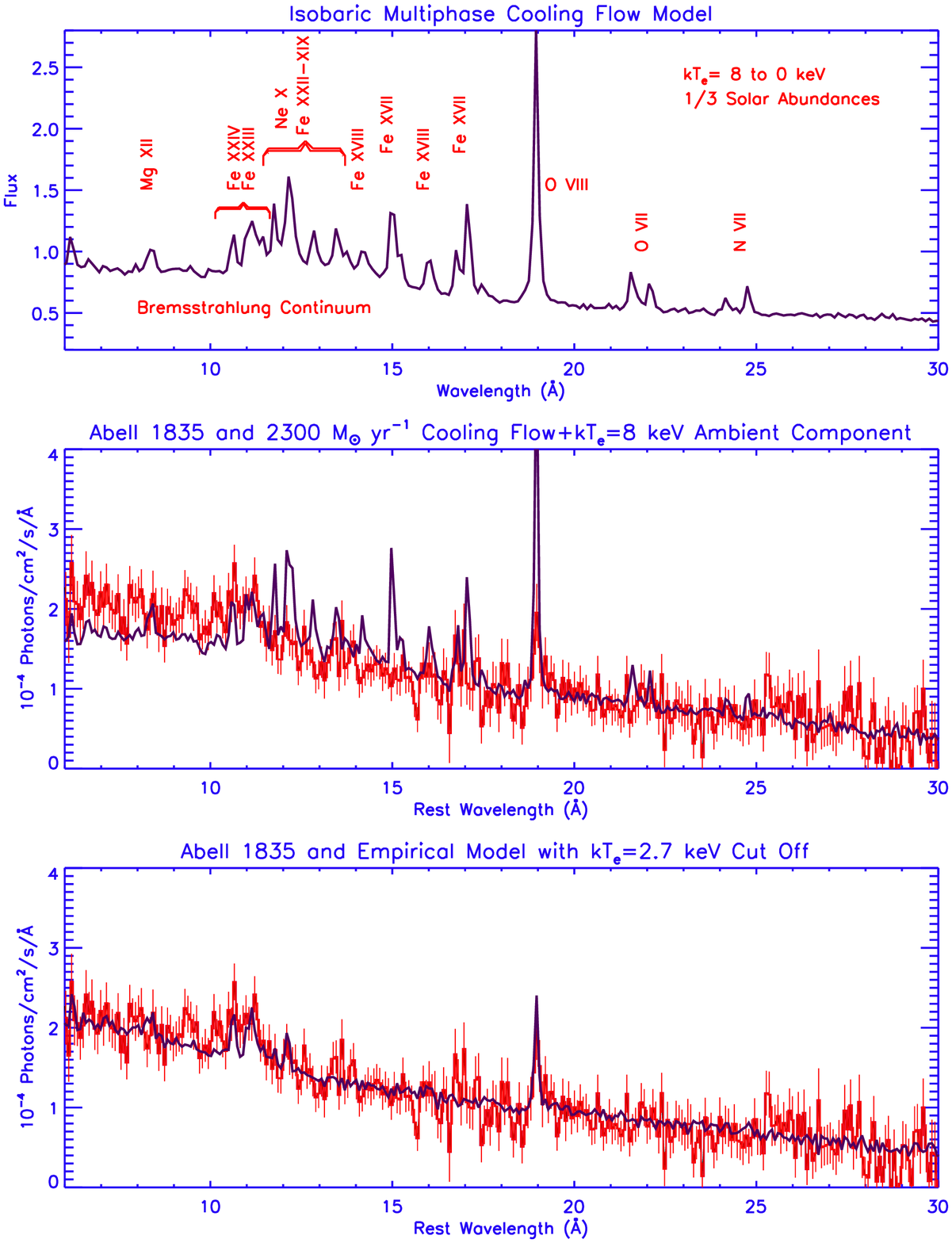} 
\caption{Left: solid squares and circles show Fe abundance in the
  inner regions of hot clusters ($kT>5$ keV) averaged over about 10
  clusters in each bin (with two different combining methods), as a
  function of redshift \cite{bal07}.  The shaded area show the rms
  dispersion of the single measures, while the dashed line is a fit to
  the $Z_{Fe}$ evolution of the form $\sim(1+z)^{-1.25}$. Right: the
  two top panels shows the model (blue) and the data from the Reflection
  Grating Spectrometer of {\sl XMM--Newton} for the cool--core cluster
  Abell 1835.  In the bottom panel the gas colder than 2.7 keV has
  been removed in the model, which now shows a good agreement with the
  data \cite{pef06}.}
    \label{fig3}
\end{figure}

We recall that the Fe line can be measured in the highest redshift
clusters selected in X--rays, as shown in the right panel of Figure
\ref{fig1}.  In a recent work, Balestra et al. (2007) used the Chandra
archive for clusters at redshift $z\geq 0.4$ to compute the average Fe
abundance in several redshift bins.  The result, shown in the left
panel of Figure \ref{fig3}, indicate that the Fe abundance in the
inner parts of X--ray hot clusters (within $\sim 0.2 R_{vir}$) is
changing by a factor of about 2 between now and $z \sim 1.3$ (which
corresponds to a look back time of $\sim 9$ Gyr).

This simple result requires a complex interpretation.  On one hand, we
notice that the Fe abundance is already significant at $z>1$, in line
with the expectation that the bulk of star formation in massive
spheroids, responsible for the large majority of the metals, occurs at
$z \geq 2$.  On the other hand, we do not expect much star formation
responsible for Iron production after $z\sim 0.5$.  Therefore, the
most likely interpretation of the increase of the average Iron
abundance in the inner $0.2 R_{vir}$ of clusters, may be due to
deposition of previously enriched gas towards the center.  Currently,
several approaches can be used.  A phenomenological model based on
detailed chemical galactic evolution, shows that the Fe increase is
consistent with the transformation of gas--rich spirals into S0
galaxies, with the consequent deposition of highly enriched gas in the
central regions where the ram pressure stripping is more
efficient \cite{cal07}.  N--body Hydrodynamic simulations, on the other
hand, show that high abundance, low entropy gas, previously associated
to galaxies or group--size halos, may sink to the center during the mass
growth of the cluster (see Cora et al. in preparation).  These models
favour a dynamical origin of the observed evolution, but they must
also explain the abundance profiles and the
temperature gradients in  cluster cores observed in spatially
resolved spectral analysis (see \cite{vik} and \cite{b07}).

\section{The nature of cool cores}

The total emission due to bremsstrahlung and lines can be expressed as
$L \propto n_e^2 \Lambda(T_{ICM}) $ where $\Lambda$ is the cooling
function including all the transitions for a given $T_{ICM}$.  The
cooling time $t_{cool}$ is defined as the ratio of the total internal
energy of the ICM divided by the bolometric ICM emission.  It turns
out that $t_{cool} \propto T_{ICM} \Lambda^{-1} n_e^{-1}$ (see
\cite{S88}).

If $t_{cool} < < t_H$ the baryons are expected to cool out of the hot
phase and eventually recombine and form stars or clouds of cold gas.
In many local clusters it is possible to compute the cooling time down
to very small scales (about few kpc).  In several cases a clear
decrease of the temperature is observed towards the center, and the
cooling time within 10 kpc can be significantly less than 1 Gyr (see
\cite{pef06}).  It seems unavoidable to predict that baryons are
flowing to the cold phase at a rate of the order of 100, sometimes
1000 $M_\odot$ yr$^{-1}$ in more than half of local clusters.  The
simplest model based on isobaric cooling, predicts a spectrum rich in
emission lines, which are strongly increasing at low $T_{ICM}$ due to
the higher number of ion species.  However, grating spectroscopy of
the brightest central regions of clusters, with the {\sl XMM--Newton}
satellite, provided a surprising result.  Many of the lines expected
in cooling flows were missing from the observed spectra \cite{p01}, as
shown in the example of Figure \ref{fig3} (right panels).  It can be
shown that the lowest temperature in the center is of the order of 1/3
of the virial one.

This discovery has a strong impact: it implies that the ICM is kept
above a temperature floor (not too far from the virial one) by some
heating mechanism. It follows that there are no more cooling {\sl
  flows}, at least not as strong as previously thought, but only
cooling cores.  On one hand, this explains why we never observed the
cooled gas resulting from the cooling process, neither in form of
stars nor of cool gas.  On the other hand, there is no consensus on
the sources which constantly heat the ICM.  This is an open problem,
relevant not only for ICM, but also for general framework of galaxy
formation and evolution

In fact, this problem reminds us of the cooling catastrophe (also
known as cooling crisis).  In the standard galactic formation
scenario, baryons in CDM halos cool via thermal bremsstrahlung and
line emission.  Baryons are assumed to turn into stars when $t_{cool}
< H^{-1}$ \cite{wr78}.  But the blind application of this criterion
would result in the large majority of the baryons locked into stars.
This is a consequence of the high power at low mass scales in the CDM
power spectrum.  The low fraction of baryons turned into stars,
observed to be around 10\% everywhere in the Universe, requires an
ubiquitous mechanism which hampers the baryons from cooling.  It would
be nice if the same heating process might explain the cooling core
problem and solve the cooling catastrophe at the same time.  This
still unknown process, or better, these class of processes, are known
under the name of {\sl feedback}, a key ingredient in every model of
cosmic structure formation.

The main problem with feedback, is that any process we may think of,
scales with volume (and then with density), while cooling is a runaway
process proportional to $\rho^2$.  Therefore there is not an obvious,
mechanism for self--regulation.  Understanding feedback is nowaday the
most compelling goal for structure formation models.  Main candidates
are SNe explosions and stellar winds (as confirmed by the presence of
heavy elements in the ICM) and the much more energetic output from
AGN.  A spectacular example that favours AGN as the best candidates as
the main heating sources, is the X--ray mage of the Persues cluster
\cite{fab06}, where hot bubbles created by the jets are pushing the
ICM apart, with a total mechanical energy sufficient to heat
significantly the diffuse baryons. Still, how and on which time scale
the energy is thermalized into the ICM is still a matter of debate.

\section{Prospects for the future and conclusions}

The results discussed so far are based on X--ray spectra from CCD,
with a modest energy resolution, and from reflection grating
spectrometers, with high energy resolution but suitable only for very
bright sources.  Given the complex spatial structure of the
thermodynamical properties of the ICM, an ideal instrument should
attain at the same time good spatial and spectral resolution.  This
can be achieved with X--ray bolometers in the soft band, where most of
the lines are.  The bolometer that is expected to be onboard of the
proposed EDGE satellite \cite{pi07} will be able to detect absorption
and emission lines from the Warm Hot Intergalactic Medium (WHIM) which
includes the majority of the baryons in the Universe \cite{co99}.
With a nominal resolution of few eV, it will be able to resolve most
of the X--ray lines, and to measure the thermal broadening and
possibly the bulk motions of the ICM in the inner regions of clusters,
directly addressing aspects like viscosity and turbulence.  In
addition, it will be possible to detect the low surface brightness ICM
in the outskirts of clusters and in filaments.

In the meanwhile, we still have to exploit the full potential of the
{\sl Chandra} and {\sl XMM--Newton} satellites, which are still
operating, and which already provided many archival cluster
observations.  The results (among many others) presented in these
Proceedings, show that line emission diagnostic in X--ray spectra of
clusters of galaxies are an unvaluable tool to study the chemical and
the thermodynamical state of the ICM.  Another aspect I wanted to
stress, is that studies of the chemical enrichment of the ICM and of
the temperature structure in cool cores are extremely important for
the entire field of galaxy and structure formation.  For example, the
mechanism responsible for the temperature floor in cool cores of
clusters may be tightly related to the one responsible for the
quenching of the star formation in massive spheroids at $z \geq 2$.
Therefore, the technological challenge toward spatially resolved, high
resolution X--ray spectroscopy, will be the way to achieve important
steps forward in many aspects of extragalactic astrophysics.

\begin{theacknowledgments}
  We would like to thank the organizers of this conference for
  providing a stimulating scientific environment.  We acknowledge
  financial contribution from contract ASI--INAF I/023/05/0 and from
  the PD51 INFN grant.
\end{theacknowledgments}

\bibliographystyle{aipproc}

\end{document}